\documentstyle[epsf,epsfig,wrapfig]{ioplppt}
\def\be{\begin{equation}}
\def\ee{\end{equation}}
\def\bea{\begin{eqnarray}}
\def\eea{\end{eqnarray}}

\begin{document}
\thispagestyle{empty}
\enlargethispage{0.5in}
\vspace{-1.cm}
\begin{flushright}
UCL/HEP 97-06
\end{flushright}
\vspace{-0.6cm}

\title{Questions in 
Two Photon Physics at LEP2; including data Monte-Carlo 
comparison~\footnote{Invited talk given
at the LEP2 Phenomenology Workshop, Oxford, U.K., 14 April 1997}.
}

\author{D.J. MILLER }

\address{University College London, Gower Street, London WC1E 6BT,
England}

\begin{abstract}A partisan review of some of the most important $\gamma
\gamma$ channels accessible at LEP 2, with special stress on the measurement of
the photon structure function $F_{2}^{\gamma}$ and on associated problems with
Monte Carlo modelling.
\end{abstract}

\section{Introduction}
 \begin{wrapfigure}[12]{r}{5.2cm}
 \vspace{-0.2cm}
 \epsfig{file=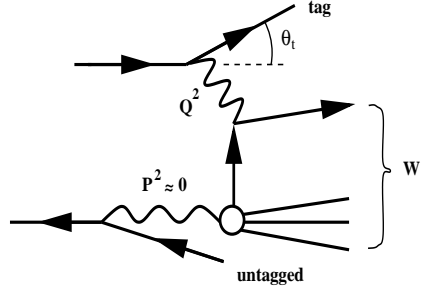,height=3.8cm,width=5.5cm}
 \vspace{-0.3cm}
 \caption{Variables in electron-photon DIS.}
 \label{fig:feynman}
 \end{wrapfigure}
There is a long agenda of possible $\gamma \gamma$ topics for LEP2 to study.
This talk dwells most upon the measurement of $F_{2}^{\gamma}$, with brief
visits to some of the rest.  I apologise in advance for OPAL-centricity.

\section{Extracting $F_{2}^{\gamma} (x,Q^{2})$}

Figure~\ref{fig:feynman} shows the Feynman graph from which the interesting
variables are defined.  The tagged lepton is either an electron or a positron,
detected in the forward or endcap detectors with a scattering angle
$\theta_{t}$ and energy $E_t$.
\[ \frac{d^2 \sigma_{e\gamma \rightarrow eX}}{dxdQ^2}= \frac{2\pi
\alpha^2}{xQ^4}[(1+(1-y)^2 )F_{2}^{\gamma}(x,Q^2)-y^2F_{L}^{\gamma}(x,Q^2)] \]
\[ Q^2=-q^2\simeq 2E_b E_t (1-cos\theta_t ) \]
\[ y=1-(E_t /E_b ) cos^2 (\theta_t /2) \]
\[ x= \frac{Q^2}{Q^2 +P^2 +W^2} \simeq \frac{Q^2}{Q^2 +W^2} \]
where $q$ is the four-momentum of the off-shell probe photon and $E_b$ is the
beam energy.  Note that
\begin{itemize}
\item It is very difficult to measure $F_L$ because the rate is lower at high y
where this term is most significant, and there are serious backgrounds
generated by beam-associated low energy electrons -- at least in OPAL and
DELPHI. (For $y>>0.5$, $E_t<<0.5$, yet most tagged studies require
$E_t>0.7E_b$.)  There is a suggestion that ALEPH may be able to do
it~\cite{Sloan} but it is hard to be hopeful.  Nothing has changed since the
1986 Aachen green book~\cite{Millergreenbook}.

\item  The double tagged rate is low with existing luminometers, but all of the
experiments are now taking their very small angle taggers more seriously so we
should soon see measurements of $P^2$ dependence from 0 to 
$\simeq 0.5 \rm ~GeV^2$.

\item  In singly tagged studies for $F_{2}^{\gamma}$ real progress has been
made, but there are practical problems, mainly because the mass $W$ of the
$\gamma \gamma$ system is hard to measure (a completely a different situation
from HERA where the target proton has a known momentum).  These 
problems will be the main focus of my talk.
\end{itemize}

\begin{figure}[t]
 \vspace{-1.3cm}
 \begin{center}
  \mbox{\hspace{-1.1cm}\epsfig{file=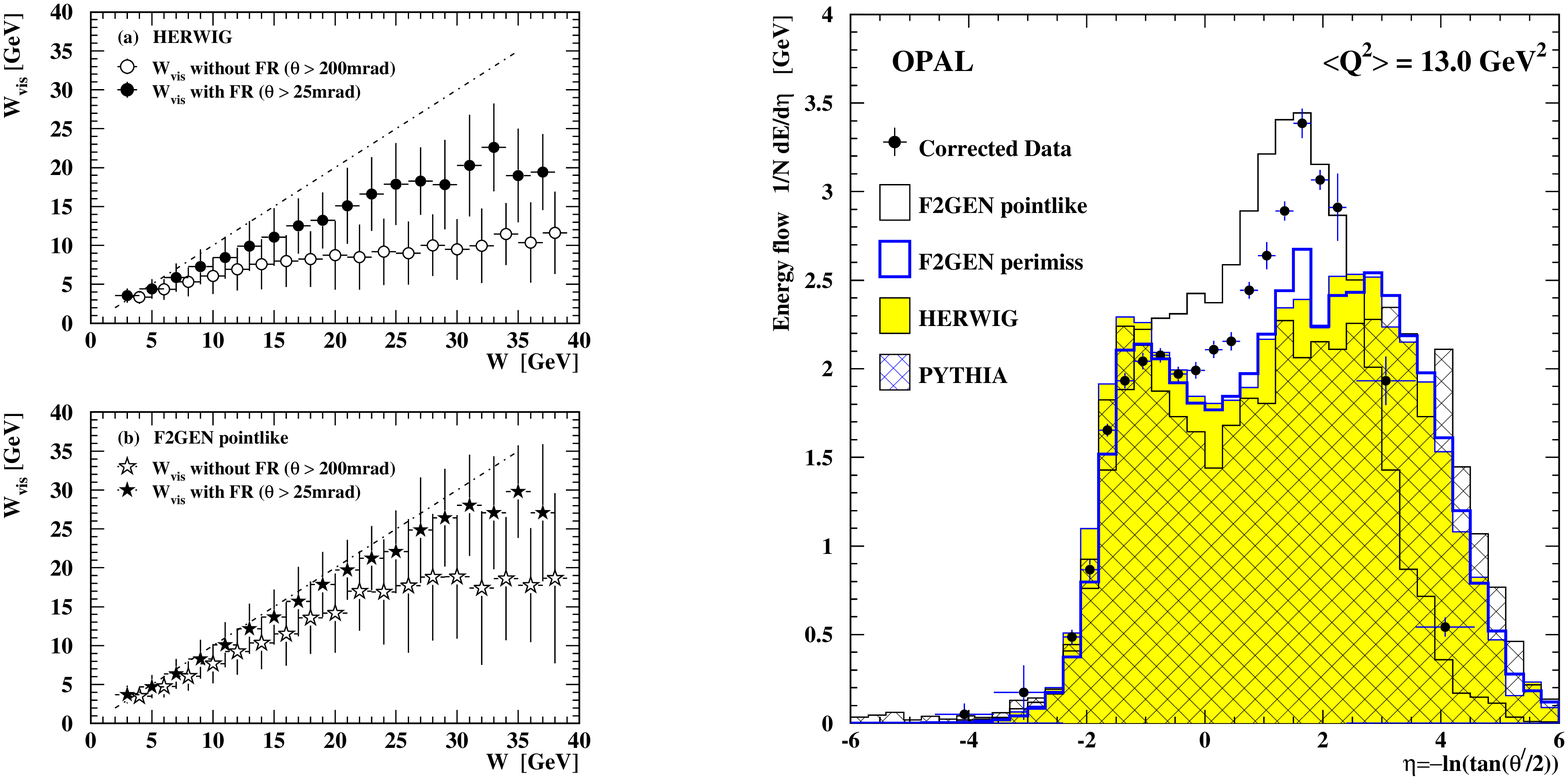,height=10cm}}

  \vspace{-2.1cm}\begin{minipage}[t]{0.47\linewidth}
   \caption{\label{fig:Wwvis} 
     $W-W_{\rm vis}$ correlation [3] for various Monte 
     Carlo models with and without the included simulation of the 
     OPAL forward region (FR) between $25<\theta <200$ mrad.}
  \end{minipage}\hfill
  \begin{minipage}[t]{0.52\linewidth}
   \caption{\label{fig:eflow}  
     Hadronic energy flow per event as a function of pseudorapidity;
     data [4] compared with various Monte Carlo models. The tag is always at
     negative $\eta$ and is not shown.}
  \end{minipage}
 \end{center}
\end{figure}

Our photons come from a very soft, spread-out spectrum.  We are forced to
measure the visible mass $W_{vis}$ of the hadronic tracks and clusters.  This
is an efficient measurement in the barrel region ($\theta >300$ mrad) but the
efficiency and accuracy both fall off in the endcaps and luminometers.  The two
standard QCD Monte Carlos, HERWIG~\cite{HERWIG} and PYTHIA~\cite{PYTHIA}, both
show a serious loss of correlation~\cite{Forshaw95} between the generated and
the visible values at large $W$ (e.g. section (a) of figure~\ref{fig:Wwvis},
where the open circles represent a measurement using only the barrel
detectors).  This lack of correlation at large $W$ corresponds to very poor
reconstruction of $x$ in the interesting region where HERA sees rising values
of the proton structure function as $x$ decreases~\cite{ZEUSfirst,H1first}.
There is great theoretical interest in knowing if the photon has a similar
behaviour to the proton at low $x$ because of its hadron-like character, or
whether the photon's 
direct coupling to quark pairs makes a visible difference.  At
first sight the HERWIG and PYTHIA predictions suggest that these measurements
might be impossible at LEP.

But there is hope.  Making use of the sampled hadronic energy in the endcaps
and luminometers, the observed energy flows seen in figure~\ref{fig:eflow} are
by no means as strongly 
forward as HERWIG or PYTHIA would suggest.  In fact, the
experimental points have some of the character of the extreme-case Monte Carlo
sample from the purely pointlike F2GEN generator~\cite{F2GEN}.  And note from
section (b) of figure~\ref{fig:Wwvis} that using this pointlike F2GEN
sample with hadronic
energy from the forward region detectors almost completely restores the
correlation  between $W_{\rm visible}$ and true $W$.  HERWIG and PYTHIA are clearly
not yet fully suitable to be used as unfolding Monte Carlo programmes for the
extraction of $F_{2}^{\gamma}$.  They must have something missing, possibly in
the way they treat hard sub-processes like photon-gluon fusion.  Interestingly,
the PHOJET~\cite{PHOJET} Monte Carlo shows signs of lying closer to the OPAL
data, in a still unpublished study with $1.5<Q^2<6 \rm ~GeV^2$, but the present
version of PHOJET is not properly set up for virtual photons.  

The working group must look at what needs to be done to produce a Monte Carlo
generator that is suitable for unfolding.  This generator should be driven by
physics from outside the $\gamma \gamma$ field but must match all aspects of
the $\gamma \gamma$ data.  The group is also in a position to study the effects
of the particular unfolding package that is used to go from the observed $x$
distribution to the true distribution. There are now two unfolding packages in
use, based on very different statistical methods, and the new ALEPH
results use
both of them: SVD~\cite{Kart} and RUN~\cite{Blobel}.  My impression is that
they agree reasonably well when the statistics are good.

The question list for the group is probably longer than the answer list will
be:
\begin{itemize}
\item  How can the QCD Monte Carlo programmes be improved?
\item  What will the remaining systematics then be?
\item  Can we ever see if there is a low $x$ rise in the structure function
like that observed for the proton at HERA?
\item  How should the charm threshold be treated: can realistic and consistent
kinematics and phase-space be used in the Monte Carlo models;  should QCD
evolution be included, or just the Quark parton Model;  what value should be
taken for the effective charm mass?
\item  Will we be able to fit for $\Lambda_{\overline{MS}}$ from the high $Q^2$
evolution?
\item  Will double tags work?
\end{itemize}

\section{Inclusive studies}

\begin{wrapfigure}[18]{r}{5.9cm}
 \vspace{0.2cm}
 \epsfig{file=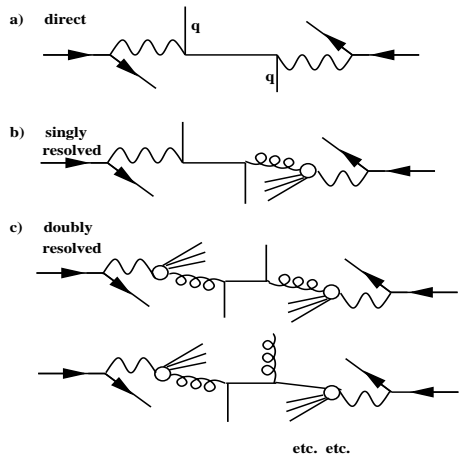,height=6.cm,width=6.cm}
 \vspace{-0.5cm}
 \caption{Feynman graphs for direct and resolved processes.}
 \label{fig:directetc}
 \end{wrapfigure}

Ever since $\gamma \gamma$ physics started at PEP and PETRA there have been
attempts to test the predictions of QCD by studying jets at high $p_T$, mostly
with untagged events.  And every experiment has developed its own tools, its
own set of variables and its own cuts~\cite{MillerCornell}.  Now OPAL has
introduced an analysis~\cite{Buergin} based very closely on what is done in
photoproduction at HERA, and there is some hope that the other LEP experiments
will follow suit.  They use a development of the $x_{\gamma}$ variable which is
an estimator of the fraction of the target photon's momentum carried by the
hard parton which produces identified jets with high $E_T$,
\[x^{\pm}_{\gamma}=\frac{\sum_{jets} {E_{j} \pm p_{z,j}}}{\sum_{hadrons}
{E_{i} \pm p_{z,i}}} \]
where $p_{z,i}$ is the momentum of the $i$th hadron projected along the beam
direction.  The $\pm$ ambiguity arises because the untagged initial state is
intrinsically symmetric, unlike the situation in $ep$, and either photon may be
the target.  Three main categories of events with high $E_T$ jets are expected:
direct, singly resolved or doubly resolved -- as shown in
Figure~\ref{fig:directetc}~\cite{ChrLls}.
Using the PYTHIA Monte Carlo, OPAL shows that the direct sample should be very
cleanly separated from the resolved samples by requiring both $x^{+}_{\gamma}$
and $x^{-}_{\gamma}$ to be greater than 0.8.  They confirm this separation in
the experimental data for two jet events with $E_{T}>3$ GeV by computing an
effective parton scattering angle $\theta^*$ in the dijet C. of M. and showing
that the ``direct'' ($x^{\pm}_{\gamma}>0.8$) sample has the  expected rather flat
distribution , while the ``resolved'' sample ($x^{+}_{\gamma}$ or $x^{-}_{\gamma}$
less than 0.8) is much more forward-backward peaked, as predicted on a parton
level by lowest order QCD (and as seen in analyses of
photoproduction at HERA~\cite{HERAjets}).  This is a new field which has opened
up at LEP 2 where the background in the untagged channel from $Z^0 \rightarrow
~hadrons$ is greatly reduced compare with LEP 1.

A development of this analysis will be to study the jet structure in tagged
events and to see how it varies as a function of the $Q^2$ of the probe photon.
It is not obvious what the equivalent variable should be in this case to $E_T$
in the untagged case, since the transverse energy in tagged events contains the
hadronic recoil against the tag.  It has been suggested that it might be
possible to use normal jet-finding algorithms in the C. of M. system of the
visible hadrons, but we know that missed forward energy will always make this a
poor approximation to the true hadronic rest frame.  Another possibility is to
use the distribution of the $E_{t,out}$ variable (see figure 1
of~\cite{Cartwrig}) instead of $E_T$.    A preliminary study by
Rooke~\cite{Rooke} has found a much larger signal in OPAL for two jet events
than the predictions of HERWIG or PYTHIA.

One of the worst measured quantities in $\gamma \gamma$ physics is the total
cross section $\sigma_{\gamma \gamma}$ for $W<5$ GeV.  L3~\cite{VanRossum} is
now producing LEP2 results at higher $W$, up to $\simeq 70$ GeV.  We should all
try to match them.  Here again, the big problem will be in correcting
believably for the lost hadrons going into the forward region and even down the
beampipe -- a significant number of diffractive events at high $W$ may give no
signal at all in our detectors.  L3 may not yet have 
understood the full extent of this problem (but see ``Utopia'', below).

\section{Resonances}
There is no doubt that we should continue to study charmonium
resonances at LEP2.
L3 has already had some success at LEP1 with $\eta_c$~\cite{Adriani} and with
$\chi_c$~\cite{Buijs}.  It is hard work because there are lots of decay modes,
but adding many channels together can give respectable peaks.  Hundreds of
events should be collected in the end for each of the charmonium resonances,
enough to give worthwhile measurements of their $\gamma \gamma$ partial widths.

It is not so clear how far we should pursue the study of
non-charmonium resonances
in the 1 to 2 GeV region.  Close and colleagues~\cite{Close} have an important
list of glueball and hybrid candidates whose ``stickiness'' $S$ has to be
checked, where for resonance $R$,
\[S=\frac{\Gamma (J/\psi \rightarrow \gamma R)/phasespace}{\Gamma^{R}_{\gamma
\gamma}/phasespace}. \]
($S \simeq 1$ for the $f^{0}_{2}(1270)$, a clear $q \overline{q}$
state which is prominent
in $\gamma \gamma \rightarrow \pi^+ \pi^-$, but $S \simeq 25$ for
$\eta (1440)$, a well known glueball candidate which is barely visible in
$\gamma \gamma$).  The integrated luminosity at LEP 2  is never expected to
exceed $500 pb^{-1}$, whereas Cleo II already has in excess of $3 fb^{-1}$ and
the new beauty factories BaBar and Belle should each get more than $10
fb^{-1}$.  Beauty factories have almost as large a production cross section for
low mass resonances as LEP, and the final states are easier to measure because
they are not so strongly boosted along the beam direction.  LEP triggers are
also heavily biased against low transverse momentum tracks and low
multiplicities.  Nevertheless, L3 is beginning to get into resonance studies.
It will be hard for the working group to judge how much should be done on
resonances at LEP 2 because no one can tell how much effort will eventually be
available for $\gamma \gamma$ resonance studies at the beauty factories.  [Note
added later.  In my judgement, the only two $\gamma \gamma
\rightarrow$ resonance results at Photon '97 which had real physics impact both
came from Cleo II, with large statistics~\cite{Paar,Savinov}.  
On one of the two topics ($Q^2$
dependence of $\eta'$ production) there was also an L3
paper~\cite{L3res} which did not
add much at this stage, though with the full LEP 2 statistics it will
eventually give a useful check].

\section{The biggest exclusive cross-sections at LEP; $\gamma \gamma
\rightarrow $vector meson pairs.}
 \begin{wrapfigure}[24]{r}{7.5cm}
 \vspace{-0.8cm}
 \epsfig{file=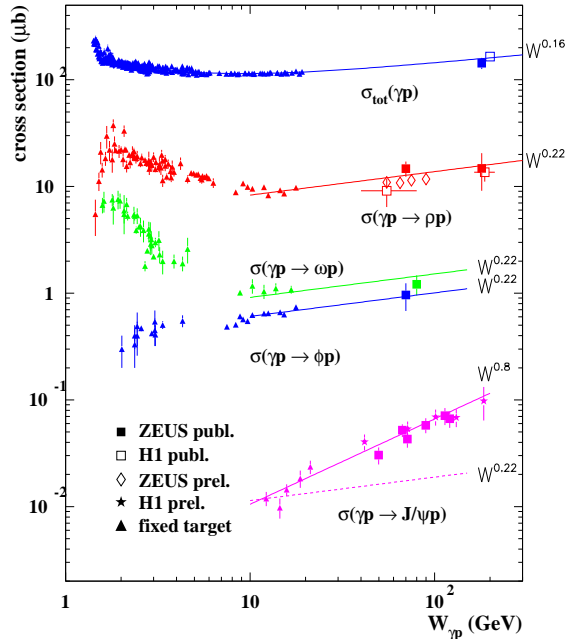,height=9.2cm,width=7.6cm}
 \vspace{-0.7cm}
 \caption{The $\gamma$p total cross section and the cross sections
  for photoproduction of single vector-mesons, including recent HERA data.}
 \label{fig:HERA}
 \end{wrapfigure}
Figure~\ref{fig:HERA} shows that the cross section for $\gamma p\rightarrow
\rho p$ at HERA remains at about 1/12 of the total $\gamma p$ cross section
from about 10 GeV to the highest visible energy.  The same kind of behaviour
is to be expected for the equivalent ``elastic'' channel $\gamma \gamma
\rightarrow \rho^0 \rho^0$.  No one has ever measured this channel for $\rho
\rho$ masses beyond  about 2.5 GeV because the process becomes almost totally
diffractive and the pions from high energy $\rho^0$ decays go strongly forward.
These events must be coming at a rate of 1 every 20 seconds or so, but it
is a real challenge to the LEP experiments to devise triggers which will
identify them, together with related channels like $\rho \omega$, $\rho \phi$
and the semi-inclusive diffractive channels with a $\rho$ plus a low mass
forward jet.  There will be good old-fashioned soft strong interaction physics
to be done with such data.  But there is even more interest in the semi-hard
processes where one or both of the vector mesons is a $J/\psi$.  Notice that
the $\gamma p \rightarrow J/\psi p$ cross section rises in Figure~\ref{fig:HERA}
much more rapidly than the
total cross section.  QCD has predictions to make about such
processes~\cite{LEPyellow}, and the final state will be much more triggerable
and measurable than $\rho \rho$ if it contains a $J/\psi \rightarrow l^+ l^-$
decay.  The $\gamma\gamma\rightarrow J/\psi\rho$
channel is well worth further study. 

\section{Utopia}
A flight of fancy: {\em if} LEP runs to the year 2000, {\em if} we had an extra
\pounds 2M to spend and {\em if} 
we could find 6 strong uncommitted groups to work
flat out for four years.  {\em Then} let us build a $0^\circ$ double-tag
spectrometer in an unused LEP straight section.  We would use the Novosibirsk
technique~\cite{Romanov}, with a simple central detector and magnetic bends
close to the mini-beta quadrupoles so that the momenta of the two tagged
electrons can be measured with very high precision by a series of small
position sensitive detectors placed along the outgoing beamlines.  Each
detector sees the image of the collision point formed by  electrons scattered
close to $0^\circ$ in a small range of momenta and angles, focussed by the
quadrupole and dispersed by the 
bending magnet.  The simplest such set-up would measure the total 
cross section with small errors up to $W_{\gamma \gamma} \simeq
100$ GeV.  Adding better large-angle taggers would 
give $F_{2}^{\gamma}(x,Q^2)$ without the current uncertainties 
in unfolding $x$.  More elaborate forward
hadron tracking would allow $\gamma \gamma \rightarrow \rho \rho$ to be done
properly.  It would be a coherent programme and not too expensive, but somehow
I do not see it happening.

\section{Conclusions and Acknowledgements}
This review is meant to be provocative rather than even-handed, and to start
the working group off with a list of issues to discuss.  It is a good time to
have a UK workshop session on the problems 
of $\gamma \gamma$ physics at LEP because a
significant number of the most active protagonists are based in this country.

The organisers of the workshop are to be congratulated on putting 
together a very sound
programme of work to be done in many fields of LEP 2 physics.  
I look forward to IOP 1/2 day meetings over
the next few years in which the LEP 2 experimental results are reported and
discussed.  

This paper owes a great deal to the scientific advice and practical
help of  Dr. Jan A. Lauber.

\vfill\eject
\end{document}